\documentclass[twocolumn, prl]{revtex4}
\usepackage{graphicx}
\usepackage{epsfig}
\usepackage{amsmath}
\usepackage{hyperref}

\begin{document}

\title{Nodal Brillouin Zone Boundary from Folding a Chern Insulator}
\author{Li-Jun Lang}
\author{Shao-Liang Zhang}
\author{Qi Zhou}
\thanks{E-mail: qizhou@phy.cuhk.edu.hk}
\affiliation{Department of Physics, The Chinese University of Hong Kong, Shatin, New
Territories, Hong Kong}
\date{\today }

\begin{abstract}
Chern insulator is a building block of many topological quantum matters,
ranging from quantum spin Hall insulators to fractional Chern insulators.
Here, we discuss a new type of insulator, which consists of two half filled ordinary Chern insulators. On the one hand,  the bulk energy spectrum is obtained from folding that of either Chern insulator. Such folding gives rise to a nodal boundary of the Brillouin zone, at which
the band crossing is protected by the symmetries of the two-dimensional lattice that is invariant under combined transformations in the spatial and the spin space. It also provides one a
natural platform to explore the non-abelian Berry curvature and the resultant quantum phenomena. On the other hand, these two underlying Chern insulators are  distinguished from each other by nonsymmorphic operators, which lead to intriguing properties absent in conventional Chern insulators. A new degree of freedom, the parity of the nonsymmorphic symmetry, needs to be introduced for describing the topological pumping, if the edge respects the nonsymmorphic symmetry .
\end{abstract}

\maketitle

In the band structure of a crystal, if different bands are separated from
each other by finite band gaps, a Chern number\cite{Thouless1982}, which is
the integral of the abelian Berry curvature in the Brillouin zone (BZ), can
be assigned to each individual band. A Chern insulator\cite{Haldane1988}
arises if filling electrons to these bands leads to a finite total Chern
number. Such Chern insulators are fundamental elements of a wide range of
topological quantum matters\cite{Hasan2010,Xiao2010,Qi2011}. For instance, one may obtain quantum spin Hall
insulators \cite{Kane2005,Bernevig2006,Konig2007} by assembling two insulators with both opposite spins and Chern numbers. Introducing interactions,
fractional Chern insulators may emerge, in analogy to fractional quantum
Hall states\cite{Regnault2011, Tsui1982,Laughlin1983,Haldane1983,Trugman1985}.

The study of topological matters using ultracold atoms has been growing fast
in the past a few years\cite{Galitski2013,Spielman2013}. Using highly controllable atomic samples, a number
of fundamentally important theoretical models have been realised, such as
the Harper-Hofstadter model with a large magnetic flux per unit cell\cite{Bloch2013,Ketterle2013} and the topological Haldane model\cite{Jotzu2014}. Meanwhile, many topological quantum
quantities or phenomena, which are difficult to trace in solids, have been
directly observed. For instance, both Zak phase and Chern
numbers have been directly measured in optical lattices\cite{Bloch2013Z,Bloch2015}.
Topological charge pumping has also been realised recently \cite%
{Bloch2016,Takahashi2016}. Whereas most of the current studies have been
focusing on abelian topological matters, it is promising that ultracold atoms
may provides physicists a platform to explore non-abelian topological
matters, as well as new quantum states and phenomena that have not been
studied in the literature.

In this Article, we study a new type of topological matter, whose bulk energy
spectrum consists of two half filled ordinary Chern insulators. The band structure can be obtained from folding of that of either Chern insulator, which has BZ doubling the one of the realistic system. As a result of the folding, energy bands form pairs and touch at
the whole zone boundary.  Moreover, such nodal boundary is protected by the
 symmetries  of the
two-dimensional lattice. One is a nonsymmorphic symmetry, a combination of
shifting the lattice by half of the lattice spacing and a
rotation of the spin.  This nonsymmorphic symmetry distinguishes the two underlying Chern insulators and gives rise to intriguing properties of the system that are absent in ordinary Chern insulators. The other is also a combination of transformations in the spatial and the spin space, a mirror reflection and corresponding flips of spin. Since the band gap closes at the
nodal boundary, non-abelian Berry curvature is required for describing the
topological properties of the band structure. This system thus allows one to
explore quantum dynamics controlled by non-abelian Berry curvature, which
has two characteristic features. First, the drift of wave packet in the
momentum space is accompanied by the change of the occupation in different
bands, which can be viewed as a rotation of a pseudospin determined by
Wilson line\cite{Wilczek1984}, the line integral of the non-abelian
connection. Second, the first Chern number, which is the integral of the
trace of the non-abelian Berry curvature, in our system is found to be 1.
The charge transferred in single pumping period is thus quantised to one,
though the contribution from each individual half-filled Chern insulator is
not quantised. Whereas the bulk properties are readily rich, we also consider edge states
of a finite system. Distinct from a conventional Chern insulator, the number
of edge states along the edges, which respect the nonsymmorphic crystalline
symmetry, depends on the momentum, since the edge BZ is also the one folded
from that of an ordinary Chern insulator. More importantly, if the edge respects nonsymmorphic crystalline symmetry, the   topological pumping acquires a new degree of freedom, the parity of the nonsymmorphic symmetry, in addition to charge.

\begin{figure}[tbhp]
\centering
\includegraphics[width=1\linewidth]{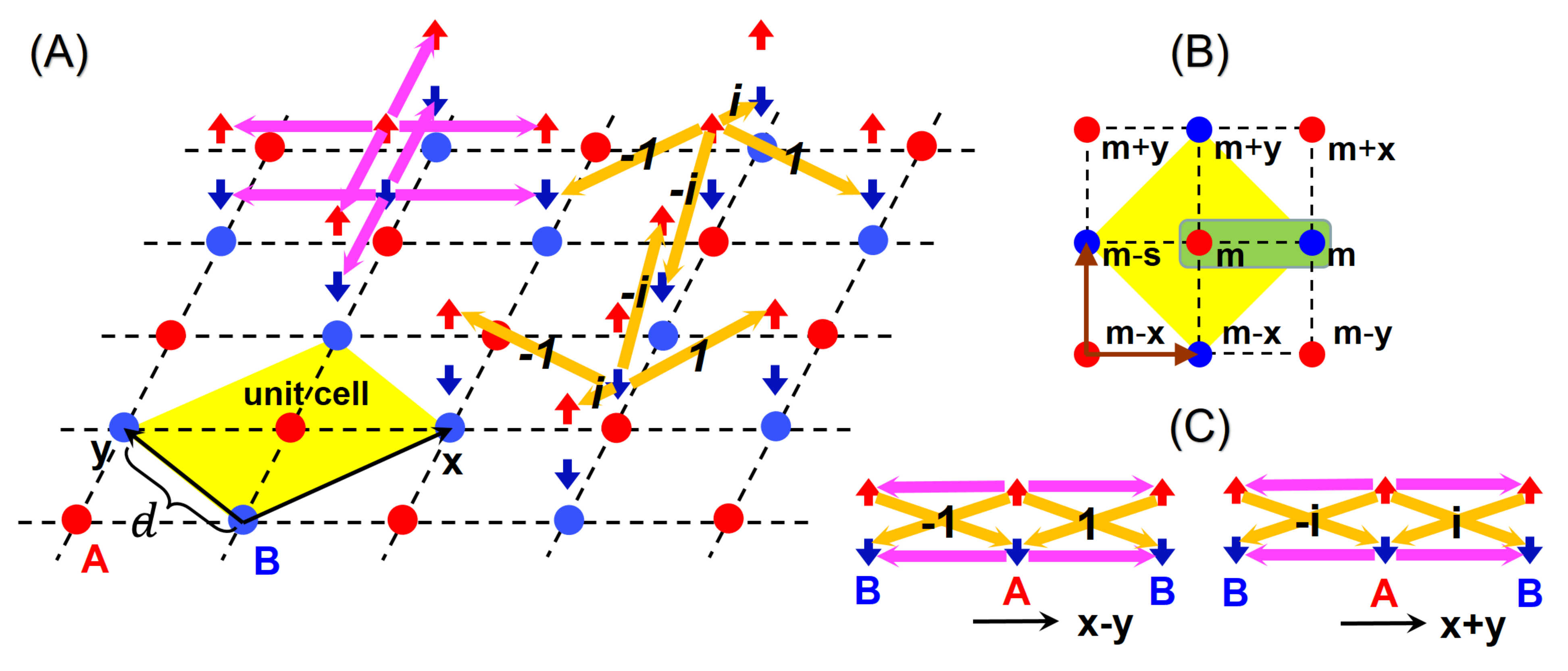}
\caption{Tight binding model. ({\bf A}) Schematics of the tunnelings. Red and blue dots represent A and B sublattice sites. Pink and orange arrows represent intra- and inter-spin tunnellings, $t$ and $t'$, respectively. $1$, $i$, $-1$, and $-i$ denote the phases of $t'$ along different directions.  A unit cell has been highlighted using yellow colour. ({\bf B})  ${\bf m}$ represents the coordinates of A sublattice sites. Its four nearest-neighbour B sublattice sites are denoted as ${\bf m}$, ${\bf m}-{\bf x}$ ${\bf m}+{\bf y}$, ${\bf m}-{\bf s}$. A basis is highlighted using green colour. Brown arrows represent the spatial part of the nonsymmorphic transformation $G_{x\pm y}$. ({\bf C}) Schematic of the tunnelling along the $\hat{x}-\hat{y}$ and the $\hat{x}+\hat{y}$ directions, respectively. The numbers $\pm 1$ and $\pm i$ represent the phase of the inter-spin tunnelling along the direction of the arrows.}
\label{fig1}
\end{figure}

\begin{figure}[tbhp]
\centering
\includegraphics[width=1\linewidth]{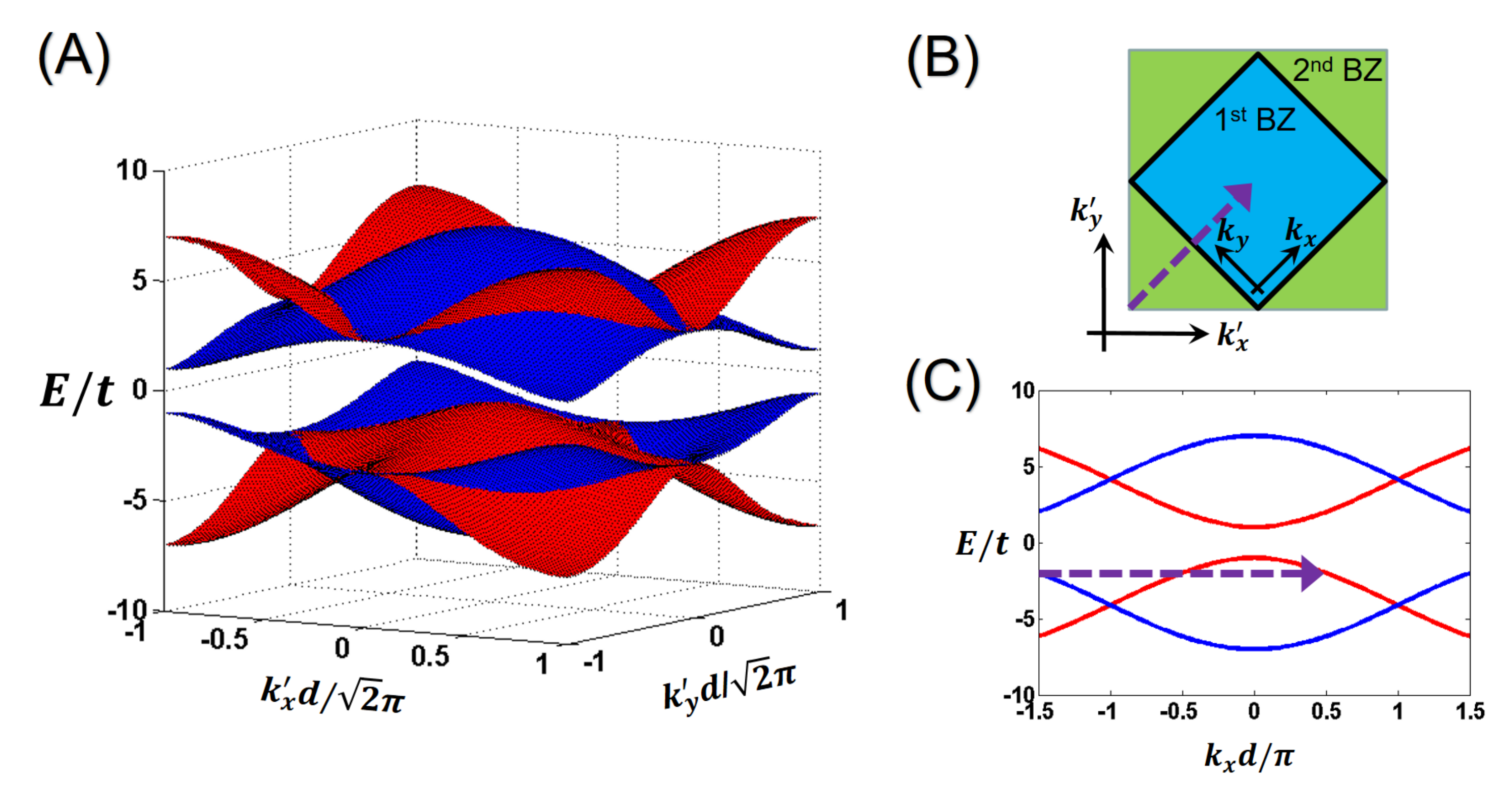}\newline
\caption{Band structure. ({\bf A}) Band structure of the Raman dressed lattice contains contributions from two Chern insulators, which correspond to $H_+(\mathbf{k})$
(red) or $H_-(\mathbf{k})$ (blue), respectively. It thus
 can be
regarded as that folding from one of these two Chern insulators. We use the tight-binding parameters as $t^{\prime}=t,
m_z=-3t.$ The lowest (highest) two bands touch at the BZ boundary, regardless of the values of $t$, $t'$ and $m_z$. ({\bf B}) BZ of the
reciprocal lattice, where the blue and green regions are the 1st and 2nd BZs, respectively.
The big square is the BZ of the Chern insulators corresponding to $H_+(\mathbf{k})$ or $%
H_-(\mathbf{k})$. The bold frame represents the nodal BZ boundary where two bands touch.  ({\bf C}) The energy spectrum along the line $k_y=0$ in (A). The purple arrows in (B) and (C) denote the reciprocal lattice vector ${\bf G}=(2\pi/d,0)$ for the folding.}
\label{fig2}
\end{figure}

\textbf{The model and its realisation}~
We consider a four-band tight binding model defined in a two-dimensional
checkerboard lattice. The Hamiltonian reads
\begin{equation}
\hat{H}%
=(\sum_{\sigma=\uparrow\downarrow}\hat{T}_\sigma)+\hat{T}_{+}+%
\hat{T}_{-}+\hat{V},\label{rtb}
\end{equation}
where $
\hat{T}_\sigma=t\sum_{\langle \mathbf{m},\mathbf{m}'\rangle}\Big(\hat{a}_{\mathbf{m}\sigma }^{\dagger
}\hat{b}_{\mathbf{m}'\sigma }+\text{H.c.}\Big)$
describes the spin-independent tunnelling $t$ between a A(B) sublattice sites and its four nearest B(A) sublattice sites. $\hat{a}_{\mathbf{m}\sigma }^{\dagger }$ ( $\hat{b}_{\mathbf{m}
\sigma }^{\dagger }$) is the creation operator for $\sigma =\uparrow
,\downarrow $ at a A (B) sublattice site ${\bf m}$.
To be explicit, $\mathbf{m}'=\mathbf{m}, \mathbf{m}-{\bf x}, \mathbf{m}+{\bf y}, \mathbf{m}-{\bf s}$, where ${\bf x}=d\hat{x}$, ${\bf y}=d\hat{y}$, and ${\bf s}={\bf x}-{\bf y}$ have been defined and $\hat{x}$($\hat{y}$) is the unit vector along the $x$($y$) direction, as shown in Fig.~\ref{fig1}.   $\hat{V}=m_{z}\sum_{\mathbf{m}}\Big(\hat{a}_{\mathbf{m}\uparrow }^{\dagger}%
\hat{a}_{\mathbf{m}\uparrow}+\hat{b}_{\mathbf{m}\uparrow }^{\dag }\hat{b}_{%
\mathbf{m}\uparrow } -\hat{a}_{\mathbf{m}\downarrow }^{\dagger}\hat{a}_{\mathbf{m}\downarrow}-%
\hat{b}_{\mathbf{m}\downarrow }^{\dag }\hat{b}_{\mathbf{m}\downarrow }\Big)$ is the Zeeman energy. The spin flip terms $\hat{T}_{\pm}$ are written as
\begin{eqnarray}
\hat{T}_{+}&=&t^{\prime }\sum_{\mathbf{m}}\Big( \hat{a}_{%
\mathbf{m}\uparrow }^{\dagger}\hat{b}_{\mathbf{m}\downarrow}-\hat{a}_{%
\mathbf{m}\uparrow}^{\dagger}\hat{b}_{\mathbf{m}-{\bf s},\downarrow}  \notag \\
&&-i\hat{a}_{\mathbf{m}\uparrow}^{\dagger}\hat{b}_{\mathbf{m}+{\bf y},\downarrow} + i\hat{a}_{\mathbf{m}\uparrow }^{\dagger}\hat{b}_{\mathbf{m}-
{\bf x},\downarrow}+\text{H.c.}\Big),\\
\hat{T}_{-}&=&t^{\prime }\sum_{\mathbf{m}}\Big( \hat{a}_{%
\mathbf{m}\downarrow }^{\dagger}\hat{b}_{\mathbf{m}\uparrow}-\hat{a}_{%
\mathbf{m}\downarrow}^{\dagger}\hat{b}_{\mathbf{m}-{\bf s},\uparrow}  \notag \\
&&+i\hat{a}_{\mathbf{m}\downarrow}^{\dagger}\hat{b}_{\mathbf{m}+{\bf y},\uparrow} - i\hat{a}_{\mathbf{m}\downarrow }^{\dagger}\hat{b}_{\mathbf{m}-
{\bf x},\uparrow}+\text{H.c.}\Big),
\end{eqnarray}
where $t'$ is the inter-spin
tunnelling amplitude. The Hamiltonian in the momentum space
is written as $\hat{H}_{\bf k}=\hat{\Psi}^\dagger_{\bf k}H({\bf k})\Psi_{\bf k}$, where $\hat{\Psi}^\dagger_{\bf k}=(\hat{a}^\dagger_{\bf k \uparrow},\hat{b}^\dagger_{\bf k\uparrow},\hat{a}^\dagger_{\bf k \downarrow},\hat{b}^\dagger_{\bf k\downarrow})$, and
\begin{equation}
H(\mathbf{k)}=\left(
\begin{array}{cccc}
m_{z} & \Omega _{\mathbf{k}}e^{i\theta_\mathbf{k}} & 0 & N_{\mathbf{k}%
}e^{i\theta_\mathbf{k}} \\
\Omega _{\mathbf{k}}e^{-i\theta_\mathbf{k}} & m_{z} & -N_{\mathbf{k}%
}e^{-i\theta_\mathbf{k}} & 0 \\
0 & -N_{\mathbf{k}}^{\ast }e^{i\theta_\mathbf{k}} & -m_{z} & \Omega _{%
\mathbf{k}}e^{i\theta_\mathbf{k}} \\
N_{\mathbf{k}}^{\ast }e^{-i\theta_\mathbf{k}} & 0 & \Omega _{\mathbf{k}%
}e^{-i\theta_\mathbf{k}} & -m_{z}%
\end{array}%
\right). \label{tb}
\end{equation}
$\Omega _{\mathbf{k}}=2t\big[ {\cos \frac{\left(k_{x}-k_{y}\right) d}{2}+\cos }\frac{\left( k_{x}+k_{y}\right) d}{2}\big]$,
$N_{\mathbf{k}}=2it^{\prime }\big[ {\sin \frac{\left( k_{x}-k_{y}\right) d}{2}-i\sin }\frac{\left( k_{x}+k_{y}\right) d}{2}\big] $, and $\theta_%
\mathbf{k}=\frac{\left( k_{y}-k_{x}\right) d}{2}$. The above Hamiltonian
can be block-diagonalised, yielding
\begin{equation}
\tilde{H}({\bf k})=\left(
\begin{array}{cccc}
m_{z}+\Omega _{\mathbf{k}} & -N_{\mathbf{k}} & 0 & 0 \\
-N_{\mathbf{k}}^{\ast } & -m_{z}-\Omega _{\mathbf{k}} & 0 & 0 \\
0 & 0 & m_{z}-\Omega _{\mathbf{k}} & N_{\mathbf{k}} \\
0 & 0 & N_{\mathbf{k}}^{\ast } & -m_{z}+\Omega _{\mathbf{k}}%
\end{array}%
\right) .  \label{block}
\end{equation}%
It is apparent that each $2\times 2$ block is a standard Hamiltonian for a
Chern insulator with Chern number $\pm 1$ if $|m_z/t|<4$. We define them as $H_{\pm }(\mathbf{k})$. The BZ of $H_{\pm }(\mathbf{k})$ doubles the one of $H(\mathbf{k})$, and
meanwhile, $H_{+}(\mathbf{k})=H_{-}(\mathbf{k+G})$, where $\mathbf{G}=(\pm \frac{2\pi }{d},0)$ or $(0,\pm \frac{2\pi }{d})$ is a reciprocal lattice vector of $H(\mathbf{k)}$. Eq.~\ref{block} tells
one that the system is composed of two Chern insulators,
and the band structure of $H(\mathbf{k)}$ can be viewed as the one folded from
of $H_{+}(\mathbf{k})$ or $H_{-}(\mathbf{k})$. Such folding can also be seen
from the real space (See the supplementary material). By a simple transformation,  $\hat{b}_{\mathbf{m}\downarrow }^{\dagger }\rightarrow -\hat{b}_{\mathbf{m}\downarrow
}^{\dagger }$, one sees that the tight binding model becomes the one for a
Chern insulator, the Hamiltonian of which in the momentum space is $H_{+}(%
\mathbf{k})$. Alternatively, the transformation $\hat{b}_{\mathbf{m}\uparrow }^{\dagger }\rightarrow -\hat{b}_{\mathbf{m}\uparrow
}^{\dagger }$ gives rise to $H_{-}(\mathbf{k})$. Since these transformations actually halve the unit cell, folding
the energy spectrum of $H_{+}(\mathbf{k})$ or $H_{-}(\mathbf{k})$ gives rise
to the band structure of the realistic system, i.e., the eigenstates of  $H(\mathbf{k)}$. In addition,  $H_{\pm}({-k_x,k_y})=SH^*_{\pm}({\bf k})S^\dag$ and $H_{\pm}({k_x,-k_y})=S^\dag H^*_{\pm}({\bf k})S$, where $S=\text{diag}(1,i)$. Thus both $H_+(\pm \frac{\pi}{d},k_y)=H_-(\pm \frac{\pi}{d},k_y)$ and $H_+(k_x,\pm \frac{\pi}{d})=H_-(k_x, \pm \frac{\pi}{d})$ are satisfied, and
band crossing occurs through the whole BZ boundary, as shown in Fig.~\ref{fig2}. At M point, ${\bf k}=(\frac{\pi}{d}, \frac{\pi}{d})$, the two linear band crossings along the $k_x$ and $k_y$ directions merge into a quadratic band touching point, i.e., the energy difference between the lowest (highest) two bands $\sim t q_xq_y$, where $q_x$ and $q_y$ are the momenta measured from the M point along $k_x$ and $k_y$ directions, respectively.  Unlike
the ordinary nodal lines or nodal rings in other systems \cite%
{Balents2011,Weng2015}, here we have nodal boundary to enclose the first BZ.
 Fig.~\ref{fig2} also shows that the lowest two bands of $H({\bf k})$ in the first BZ are contributed by $H_{+}({\bf k})$ and $H_{-}({\bf k})$, respectively. For a band insulator with one particle per spin per unit cell, either the lowest band of $H_{+}(\mathbf{k})$ or $H_{-}(\mathbf{k})$ is only half filled.  This insulator is thus formed by assembling two half-filled Chern insulators in the momentum space in a unique means. As discussed later, these two insulators are distinguished from each other by a nonsymmorphic symmetry, which leads
to a variety of unique properties of the systems in both the momentum and the real space.

The model shown in Eq.~\ref{rtb} can be realised using a number of schemes in
ultracold atoms. In the main text, we focus on one scheme that has been
realised in current experiment\cite{Wu2015}, the corresponding Hamiltonian
in the real space is written as
\begin{equation}
H(\mathbf{r})=\bigg[ \frac{\mathbf{p}^{2}}{2M}+V(\mathbf{r})\bigg] \sigma
_{0}+\Omega _{x}(\mathbf{r})\sigma _{x}+\Omega _{y}(\mathbf{r})\sigma
_{y}+m_{z}\sigma _{z},  \label{rH}
\end{equation}%
where $V(\mathbf{r})=V\big[ \cos ^{2}\frac{\pi \left(
x-y\right) }{d}+\cos ^{2}\frac{\pi \left( x+y\right) }{d}\big] $ describes
an ordinary square lattice with lattice spacing $d/\sqrt{2}$ for spin-up and
spin-down fermions, and $\Omega _{x}(\mathbf{r})=\Omega \cos \frac{\pi
\left( x-y\right) }{d}\sin \frac{\pi \left( x+y\right) }{d},\Omega _{y}(%
\mathbf{r})=-\Omega \sin \frac{\pi \left( x-y\right) }{d}\cos \frac{\pi
\left( x+y\right) }{d}$ with $\Omega $ representing the Raman coupling
strength. A finite $\Omega$ doubles the unit cell. We find that in the limit where the Raman coupling strength is much weaker than the band gap between the $s$ and $p$ bands of the square lattice, the exact band structure and topological properties of the system are well captured by the tight-binding model described before. Eq.~\ref{rH} was recently produced in reference \cite{Wu2015}.
However, reference \cite{Wu2015} just explored half of the four bands in the first BZ. Thus, the band crossing
at the zone boundary and the resultant non-abelian topological properties
were not studied. Also, to measure the Chern number, a proper method based on non-abelian Berry curvature should be used, as discussed below. An alternative experimental scheme for
realising the same nodal boundary in the momentum space is presented in the supplementary material.

The existence of the nodal boundary can also been understood from the
symmetry. We define a nonsymmorphic operator $G_{x-y}=T_{\hat{x}}(\frac{d}{2})T_{\hat{y}}(-\frac{d}{2})R_{\frac{\pi}{2}}$,  where $T_{\hat{x}}(l)$($T_{\hat{y}}(l))$ is the
translation along the $x$($y$) directions for a distance $l$, and
$R_{\frac{\pi}{2}}=e^{-i\frac{\pi}{2}\sigma_z}$ is spin rotation about the z axis for $\pi$. Since under the transformation $G_{x-y}$: $x\rightarrow x+d/2$, $y\rightarrow y-d/2$, $\sigma _{x}\rightarrow -\sigma _{x},\sigma _{y}\rightarrow -\sigma
_{y}$, one observes that $[H,G_{x-y}]=0$.  As $G_{x-y}^{2}=T_{\hat{x}}(d)T_{\hat{y}}(d)R_{\pi}$ has the eigenvalue $-e^{i(k_x-k_y)d}$, where the minus sign comes from rotating a spin-1/2 for $2\pi$, one conclude that $G_{x-y}\Psi _{n\mathbf{k}}(\mathbf{r})=\pm ie^{i\left(
k_{x}-k_{y}\right) d/2}\Psi _{n\mathbf{k}}(\mathbf{r})$, where $\Psi _{n\mathbf{k}}(\mathbf{r})$ is the Bloch function with band index $n$ and quasi-momentum $\mathbf{k}$. Similarly, we obtain $[H,G_{x+y}]=0$ and $G_{x+y}\Psi _{n\mathbf{k}}(\mathbf{r})=\pm ie^{i\left(
k_{x}+k_{y}\right) d/2}\Psi _{n\mathbf{k}}(\mathbf{r})$, where $G_{x+y}=T_{\hat{x}}(\frac{d}{2})T_{\hat{y}}(\frac{d}{2})R_{\frac{\pi}{2}}$. This is not surprising, since $G_{x+y}=G_{x-y}T_{\hat{y}}(d)$, and $H$ is commutable with both $ T_{\hat{y}}(d)$ and $G_{x-y}(d)$. Nonsymmorphic crystalline symmetry has
been considered in a variety of systems\cite{Shiozaki2015,Fang2015,Wang2016,Zhou2016}. Here, we have the nonsymmorphic crystalline symmetry along both directions in this two-dimensional
lattice. As the momentum changes by $2\pi /d$ along either the $k_{x}$ or $k_{y}$ direction, $+$ and $-$ in front of the eigenvalues of $G_{x-y}$ or $G_{x+y}$ must switch with each other, one conclude that there
must be band crossing in BZ. Moreover, there are additional symmetries along
$x$ and $y$ directions, which are represented by $M_{x}: x\rightarrow-x, \sigma_x\leftrightarrow\sigma_y$ and $M_{y}: y\rightarrow -y, \sigma_x \leftrightarrow -\sigma_y $,
respectively. Following the argument in reference\cite{Wang2016}, one
concludes that such band crossing must occur at the zone boundary, as $M_{y}G_{x-y}=T_{\hat{y}}(d)G_{x-y}M_{y}$ ($M_{x}G_{x+y}=T_{\hat{
x}}(d)G_{x+y}M_{x}$) so that $G_{x-y}$($G_{x+y}$) and $M_{x}$($M_{y}$) are
anti-commutable at $k_{x}=\pm \pi /d$ ($k_{y}=\pm \pi /d$). The system thus
has a double degeneracy at the zone boundary, i.e., a nodal boundary to
enclose the first BZ.

Using $\lambda_{x\pm y}$ to denote the sign of the eigenvalues of operators $G_{x\pm y}$, one sees that $\lambda_{x+y}=\lambda_{x-y}$,  as $G_{x+y}=G_{x-y}T_{\hat{y}}(d)$. We thus define $\lambda=\lambda_{x+y}=\lambda_{x-y}$, which is referred to as the parity of the nonsymmorphic symmetry. The wave functions of ground and first excited bands, $\Psi_{1\mathbf{k}}(\mathbf{r})$ and $\Psi_{2{\bf k}}(\mathbf{r})$,
have $\lambda =+$ and $-$, respectively, in the case of $m_z/t>0$. It is worth
mentioning that the difference of $\lambda$ between these two bands is measurable (Supplementary Materials).

\textbf{Non-abelian Berry curvature and Wilson loop}
Due to the band crossing at the zone boundary, the abelian Berry curvature is no longer
capable for describing the topological properties of the system, and
non-abelian Berry curvature is inevitably required. Under the condition that $w\ll\Delta $, where  $w$ is the total band width
of the lowest two bands and $\Delta $ is band gap between the second
and the third bands, the lowest two bands are nearly degenerate and the highest two ones can be ignored. Non-abelian Berry
curvature is defined as\cite{Xiao2010}
\begin{equation}
\mathbf{F}=\nabla _{\mathbf{k}}\times \mathbf{A}-i\mathbf{A}\times \mathbf{A}%
,
\end{equation}%
where $\mathbf{A}^{nm}=i\left\langle u_{\mathbf{k}%
}^{n}\right\vert \nabla _{\mathbf{k}}\left\vert u_{\mathbf{k}%
}^{m}\right\rangle $ is the non-abelian Berry connection, and $\left\vert u_{%
\mathbf{k}}^{n}\right\rangle $ is the periodic Bloch function with
band index $n$. It determines the dynamics of wave packets uploaded to the
system \cite{Niu2005}. Preparing an atomic cloud, say, a Bose-Einstein
condensate at an initial momentum state $\mathbf{k}_{i}$, and applying an
external force, $\mathbf{F}_{\text{ext}}$, whose strength
satisfies $w\ll |\mathbf{F}_{\text{ext}}|d\ll \Delta $, the wave packet
dynamics is dominated by the lowest two bands. We define the wave packet as $
\left\vert w\right\rangle =\sum_{\mathbf{k}}\left( \alpha _{\mathbf{k}%
}\left\vert \psi_{\mathbf{k}}^{1}\right\rangle +\beta _{\mathbf{k}}\left\vert
\psi_{\mathbf{k}}^{2}\right\rangle \right)$,
where $|\psi_{\mathbf{k}}^n\rangle$ is the Bloch function for $n$-th band, and $\eta_{\mathbf{k}} =(\alpha_\mathbf{k} ,\beta_\mathbf{k} )^T$ with $|\alpha_\mathbf{k}|^2 +|\beta_\mathbf{k}|^2=1$ may be
regarded as a pseudospin index. To simplify the notations, we have assumed
a delta function wave packet in the momentum space. The semiclassical
dynamics is described by equations \cite{Niu2005},
\begin{eqnarray}
\hbar \dot{\mathbf{k}} &=&\mathbf{F}_{\text{ext}}, \\
\hbar \dot{\mathbf{r}} &=&\nabla _{\mathbf{k}}\langle w|H_\mathbf{k}%
|w\rangle -\hbar \dot{\mathbf{k}}\times\eta_{\mathbf{k}}^{\dag }\mathbf{F}\eta_{\mathbf{k}},
\label{sc-real} \\
i\hbar \dot{\eta}_{\mathbf{k}} &=&-\hbar \dot{\mathbf{k}}\cdot \mathbf{A}\eta_{\mathbf{k}},
\label{sc-spin}
\end{eqnarray}%
which shows that the movement in the momentum space is accompanied by the
change of $\eta_{\mathbf{k}} $, i.e., the rotation of the pseudospin reflecting the
non-abelian nature of the dynamics.

\begin{figure}[tbhp]
\centering
\includegraphics[width=1\linewidth]{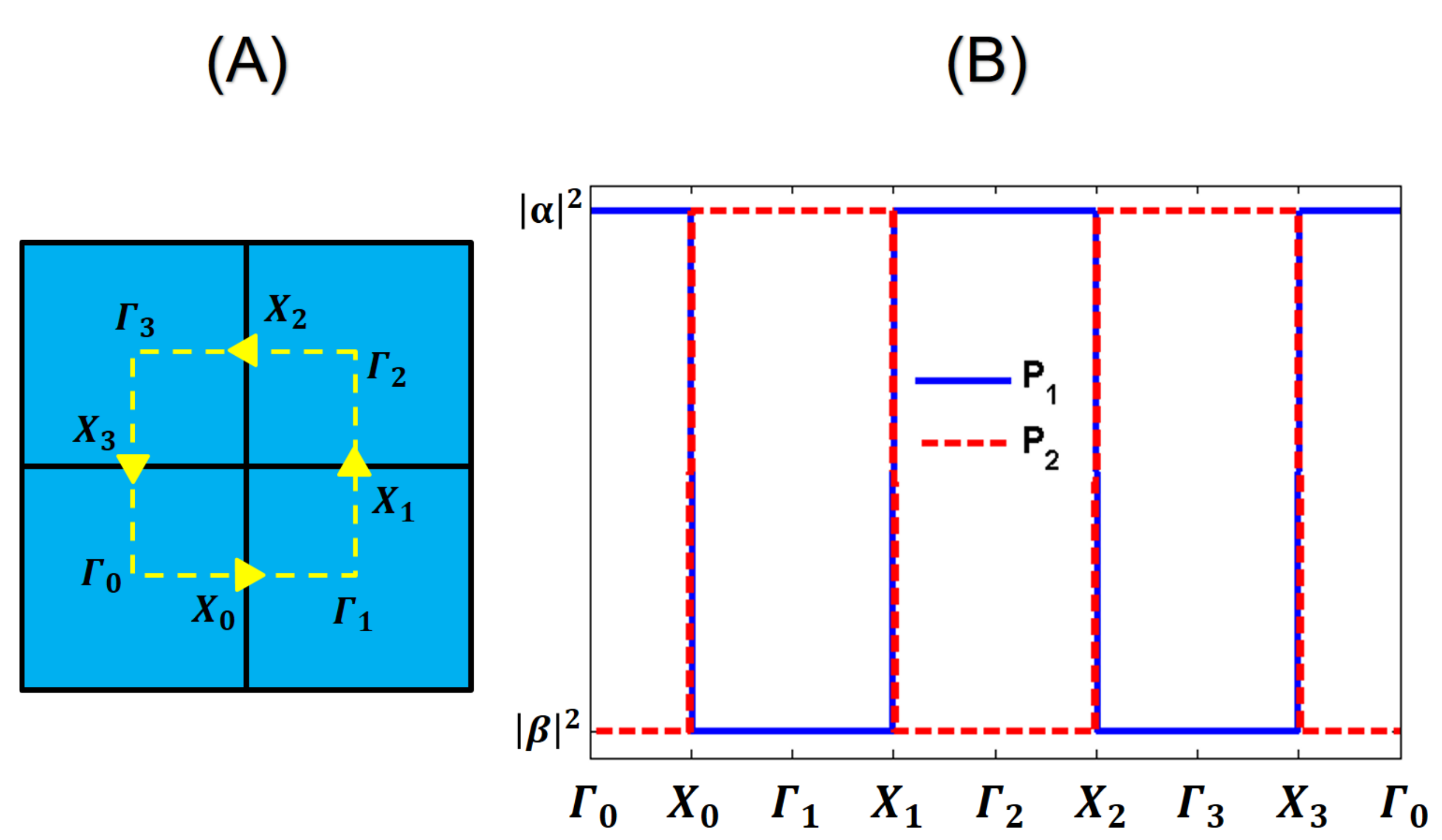}\newline
\caption{Wilson loop. ({\bf A})  The oriented yellow dashed line represents a loop in the momentum space, which intersects with the nodal boundary (bold black lines) for four times. $\Gamma_0$ is the centre of the 1st BZ, and $\Gamma_{1,2,3}$ are its equivalent points. ({\bf B}) Probabilities of the final state  on the first and second bands, $P_1$ (blue solid) and $P_2$ (red dashed), if the initial state is prepared as $\eta_{\mathbf{k}_i}=(\alpha,\beta)^T$. The parameters of the tight-binding model are the same to those in Fig.~\ref{fig2}.}
\label{fig3}
\end{figure}

Eq.~\ref{sc-spin} can be written as $
\eta_{\mathbf{k}_{f}} =W_{\mathbf{k}_{i}\rightarrow \mathbf{k}_{f}}\eta_{\mathbf{k}_{i}}$,
where $W_{\mathbf{k}_{i}\rightarrow \mathbf{k}_{f}}=\mathcal{P}e^{i\int_{\mathbf{k}%
_{i}}^{\mathbf{k}_{f}}\mathbf{A}\cdot d\mathbf{k}} $
is the Wilson line\cite{Wilczek1984}, and $\mathcal{P}$ is the path-ordering
operator. Tracing $\eta_{\mathbf{k}}$ thus allows one to reconstruct Wilson line, as
shown in a recent experiment \cite{Schneider2015}. One could also explore
the Wilson loop of this system, which is defined as
\begin{equation}
W=\mathcal{P}e^{i\oint \mathbf{A}\cdot d\boldsymbol{k}}.
\end{equation}%
For a Wilson loop shown in Fig.~\ref{fig3}, if one chooses $\eta_{{\bf k}_i}=(1,0)^T$, $|W_{11}|^2$ and $|W_{21}|^2$ correspond to the probabilities in the first and second bands, respectively. The step function-like $|W_{11}|^2$ and $|W_{21}|^2$ reflect the fact that $\lambda$ is conserved in the evolution, and the lowest two bands cross with each other at the BZ boundary\cite{Zhou2016}. We find out that the explicit form of the Wilson loop in Fig.~\ref{fig3}A is written as
 \begin{equation}
W=\left(
\begin{array}{cc}
 e^{i\varphi_{-}} & 0 \\
0 & e^{i\varphi_{+}} %
\end{array}%
\right)
,~~~~\varphi_{+}+\varphi_{-}=\int_S\text{tr}({\bf F})\cdot d\mathbf{S}. \label{Wl}
\end{equation}%
where $\int_S$ is the integral within the area enclosed by the Wilson loop. Though $W$ is gauge dependent, and $W\rightarrow U^\dagger WU$ after a local transformation of the basis, if $S$ is the whole first BZ, its U(1) part, $e^{i(\varphi_{+} +\varphi_{-})/2}$ remains unchanged and is given by the first Chern number, as shown later.

\textbf{Chern number and anomalous velocity}~
The last term in Eq.~\ref{sc-real} describes the anomalous velocity \cite%
{Niu2005,Xiao2010}. Considering an insulator with fermions fully filling up
the lowest two bands, the average anomalous velocity per particle is given
by the integral of the trace of the non-abelian Berry curvature, $\mathbf{v}_{a}=-\frac{d^2}{4\pi h}\mathbf{F}_\text{ext}\times\int_{BZ}\text{tr}( \mathbf{F})dk_xdk_y$.
Applying a dragging force, $\mathbf{F}_\text{ext}=f\hat{y}$, to the atomic cloud along the $y$ direction, one gets the transverse velocity,
\begin{equation}
{\bf v}_a= -\frac{d^2}{2 h}f C \hat{x}\equiv -v_{ax}\hat{x}\label{vac}
\end{equation}
where
\begin{equation}
C=\frac{1}{2\pi }\int_{BZ}\text{tr}\left( F_{z}\right) dk_{x}dk_{y},
\end{equation}%
is the first Chern number of the non-abelian system. Eq.~\ref{vac} allows one to
experimentally measure the Chern number using the anomalous velocity along
the transverse direction when dragging an atomic cloud, similar to the
abelian case studied in a recent experiment\cite{Bloch2015}. Using the particle density $\rho=2/d^2$, one could also obtain an effective Hall conductance, $\sigma_{xy}=j_x/E_y=q\rho v_{ax}/(f/q)=Cq^2/h$, where the charge $q$ for neutral atoms can be set as 1 or replaced by the mass $m$ for describing mass current.

Using the exact solution of the Hamiltonian, $C$ in our system is found to
be 1. Such Chern number can also be understood from the block-diagonalised
Hamiltonian in Eq.~\ref{block}. As discussed before, such equation tells
one that our system can be viewed as a composition of two Chern insulators
governed by $H_{+}(\mathbf{k})$ and $H_{-}(\mathbf{k})$, respectively. One
thus compute the Chern number of each of them using the corresponding periodic Bloch functions, $|u_{\mathbf{k}}^\pm\rangle$. It is obtained from
integrating the abelian Berry curvatures $\mathbf{F}^{\pm
}=\nabla _{\mathbf{k}}\times \mathbf{A}^{\mathbf{\pm }}$ with ${\mathbf{A}%
^{\pm }=i\left\langle u_{\mathbf{k}}^{\pm }\right\vert \nabla _{\mathbf{k}%
}\left\vert u_{\mathbf{k}}^{\pm }\right\rangle }$ as
\begin{equation}
C_{\pm }^{\prime }=\frac{1}{2\pi}\int_{BZ^{\prime }}F_{z}^{\pm }dk_{x}dk_{y}=1,
\label{abelianChern}
\end{equation}
where BZ${}^{\prime }$ denotes the BZ of $H_{+}(\mathbf{k})$ and $%
H_{-}(\mathbf{k})$. Since BZ${}^{\prime }$ doubles BZ of the system, i.e.,
the one of $H(\mathbf{k})$, Chern number of either of them is not well
defined in the first BZ of the system. Nevertheless, because of the relation
$H_{+}(\mathbf{k})=H_{-}(\mathbf{k+G})$, one obtains
\begin{equation}
C_{+g}+C_{-g}=1,
\end{equation}%
where the subscript $g$ denotes the ground bands of $H_{+}$ and $H_{-}$, and
\begin{equation}
C_{\pm g}=\frac{1}{2\pi}\int_{BZ}F_{z}^{\pm }dk_{x}dk_{y}.
\end{equation}
This observation tells one that, though the Chern number, which controls the contribution to the charge
pumping from each individual Chern insulator is not quantised in a single
pumping period, the sum of their contributions gives a well defined quantised
Chern number and thus charge pumping. This picture is also useful for understanding the Wilson loop in Eq.~\ref{Wl}. As $\lambda=\pm$ is conserved along the loop, in the basis of eigenstates of $H_{\pm}({\bf k})$, Wilson loop is diagonal, and the phase accumulated is given by $2\pi C_{\pm g}$, the sum of which is indeed the trace of ${F}_z$.

If one could selectively populate different bands, even more
interesting phenomena may occur. Instead of filling up the lowest two bands,
now consider filling up the first and third bands, which corresponds to the
ground and excited bands of $H_{+}$ and $H_{-}$, respectively. Since the
Chern number changes sign in the excited band, one sees that the net charge
pumped, which is given by $C_{+g}+C_{-e}=C_{+g}-C_{-g}$, is not quantised in
a single period. The charge currents from these two insulators apparently have
opposite directions. On the other hand, one note that $H_{+}$ and $H_{-}$
correspond to $\lambda =+$ and $\lambda =-$ respectively, one could define a
parity or pseudospin current as
\begin{equation}
j_{s}\equiv j_{+}-j_{-}=\left( C_{+g}-C_{-e}\right) \frac{q^{2}E_y}{h}=\frac{q^{2}E_y}{h}.
\end{equation}%
One thus has a quantised pseudospin pumping.

\begin{figure}[tbph]
\centering
\includegraphics[width=1\linewidth]{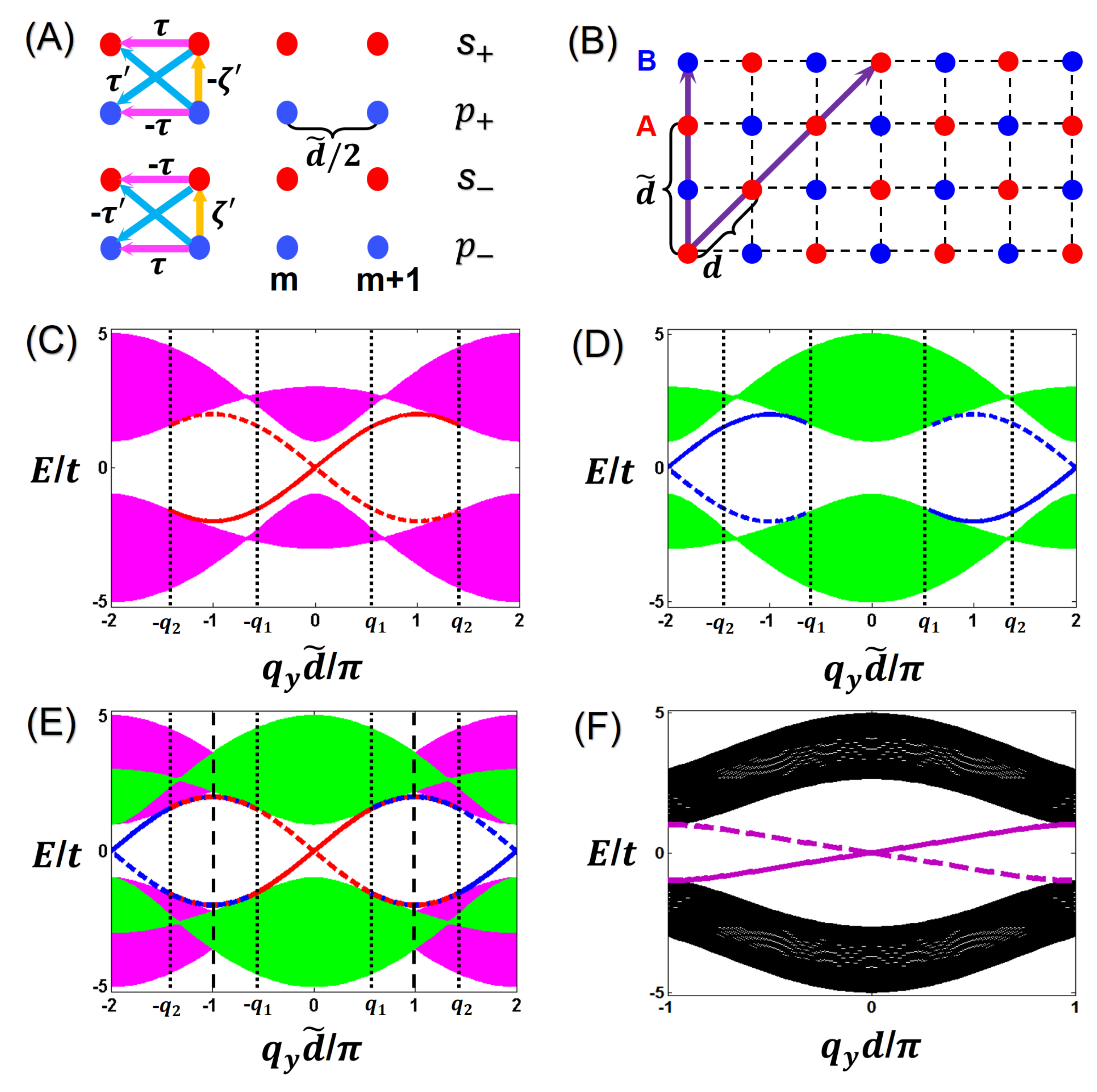}
\caption{({\bf A}) Schematics of the effective one dimensional
$s$-$p$ models with lattice spacing $\tilde{d}/2$ and the effective tunnellings labeled in the
figure, which are defined in the text. ({\bf B}) Two different boundaries, A-B-A and A-A-A, denoted by the purple arrows, which have the lattice spacings $\tilde{d}$ and $d$, respectively. ({\bf C,D}) Energy spectra with open
boundaries for block-diagonalised Hamiltonians, $\hat{h}_+(q_y)$ and
$\hat{h}_-(q_y)$, respectively. ({\bf E}) Realistic spectra with
A-B-A boundaries with period $2\pi/\tilde{d}$ (the region between
the two black dashed vertical lines), which can be regarded as folding
from one of the effective one-dimensional $s$-$p$ models (with
corresponding colors in (C) and (D)) with double period. In the region of $q_1\pi\leq |q_y|\tilde{d}\leq q_2\pi$ between two dotted vertical lines, the edge states are doubled in each boundary. ({\bf F})
Energy spectrum with A-A-A boundaries as comparison. The solid
(dashed) purple lines represent the edge states on the right (left)
boundaries. Here we use $t'=t$ and $m_z =-t$.}
\label{fig4}
\end{figure}

\textbf{Edge states}~
In a Chern insulator, it is well known that chiral edge states exist in a
finite system, and the number of edge states is directly controlled by the
Chern number of the bulk. Here, we consider the $A-B-A$
edge of our system, which respects the nonsymmorphic crystalline symmetry of
the bulk. Since the crystal momentum along the edge, which is defined as $%
q_{y}$, is a good quantum number, one considers a one-dimensional problem
for each fix $q_{y}$, and the tight binding model for such one-dimensional
system is written as $\hat{h}(q_{y})=(\sum_{\sigma=\uparrow%
\downarrow}\hat{t}_\sigma)+\hat{t}_{+}+\hat{t}_{-}+\hat{v}$, where $\hat{v}= m_{z}\sum_{m}\Big( \hat{a}_{m\uparrow }^{\dagger } \hat{a}
_{m\uparrow }+\hat{b}_{m\uparrow }^{\dagger }\hat{b}_{m\uparrow }-\hat{a}_{m\downarrow }^{\dagger }\hat{a}_{m\downarrow }
-\hat{b}_{m\downarrow }^{\dagger }\hat{b}_{m\downarrow }\Big)$ is the Zeeman energy,
\begin{eqnarray}
\hat{t}_\sigma &=&t\sum_{m}\Big[\left( 1+e^{-iq_{y}\tilde{d}}\right) \hat{a}%
_{m\sigma }^{\dagger }\hat{b}_{m\sigma }+\hat{a}_{m\sigma }^{\dagger }\hat{b}%
_{m-1,\sigma }  \notag \\
&&+e^{-i q_{y}\tilde{d}}\hat{a}_{m\sigma }^{\dagger }\hat{b}_{m+1,\sigma }+%
\text{H.c.}\Big] \\
\hat{t}_{+}&=&t^{\prime }\sum_{m}\Big[\left( 1-e^{-i%
q_{y}\tilde{d}}\right) \hat{a}_{m\uparrow }^{\dagger }\hat{b}_{m\downarrow}+ i \Big(%
\hat{a}_{m\uparrow }^{\dagger }\hat{b}_{m-1,\downarrow }  \notag \\
&&-e^{-iq_{y}\tilde{d}}\hat{a}_{m\uparrow }^{\dagger }\hat{b}%
_{m+1,\downarrow }\Big)+\text{H.c.}\Big]\\
\hat{t}_{-}&=&t^{\prime }\sum_{m}\Big[\left( 1-e^{-i%
q_{y}\tilde{d}}\right) \hat{a}_{m\downarrow }^{\dagger }\hat{b}_{m\uparrow}- i \Big(%
\hat{a}_{m\downarrow }^{\dagger }\hat{b}_{m-1,\uparrow }  \notag \\
&&-e^{-iq_{y}\tilde{d}}\hat{a}_{m\downarrow }^{\dagger }\hat{b}%
_{m+1,\uparrow }\Big)+\text{H.c.}\Big]
\end{eqnarray}
and $\tilde{d}=\sqrt{2}d$ is the periodicity of such an effective one-dimensional model along the $A-B-A$ direction. This Hamiltonian can also be block-diagonalised as $\hat{h}(q_{y})=\sum_{\lambda=\pm}\hat{h}_\lambda(q_{y})=\sum_{\lambda=\pm}\left(%
\hat{t}_{\lambda}+\hat{e}_{\lambda}+\hat{v}_\lambda\right)$, with
\begin{eqnarray}
\hat{t}_{\lambda}&=&\lambda \sum_{m}\left[ \tau\left( \hat{s}_{m\lambda
}^{\dagger }\hat{s}_{m+1\lambda }-\hat{p}_{m\lambda }^{\dag }\hat{p}%
_{m+1\lambda }\right)+\text{H.c.}\right] \\
\hat{e}_{\lambda}&=&\lambda \sum_{m}\Big[\tau^{\prime }\left(\hat{s}%
_{m\lambda }^{\dagger }\hat{p}_{m+1\lambda }+\hat{p}_{m\lambda }^{\dagger }%
\hat{s}_{m+1\lambda }\right)  \notag \\
&&-\zeta^{\prime }\hat{s}_{m\lambda }^{\dagger }\hat{p}_{m\lambda }+\text{%
H.c.}\Big], \\
\hat{v}_\lambda &=&\left( m_{z}+\lambda \zeta\right)\sum_{m} \left( \hat{s}%
_{m\lambda }^{\dagger }\hat{s}_{m\lambda }-\hat{p}_{m\lambda }^{\dagger }%
\hat{p}_{m\lambda }\right),
\end{eqnarray}%
where the new basis is $\hat{s}_{m\pm }=\left( \hat{a}%
_{m\uparrow }\pm e^{-iq_{y}\tilde{d}/{2}}\hat{b}_{m\uparrow }\right) /\sqrt{2}$
and $\hat{p}_{m\pm }=\left( \hat{a}_{m\downarrow }\mp e^{-iq_{y}\tilde{d}/{2}}%
\hat{b}_{m\downarrow }\right) /\sqrt{2}$, and the redefined parameters are $%
\tau=te^{-iq_y\tilde{d}/{2}}$, $\tau^{\prime }=it'e^{-iq_y\tilde{d}/{2}}$, $%
\zeta=2t\cos\frac{q_y\tilde{d}}{{2}}$, and $\zeta^{\prime }=2it^{\prime }\sin%
\frac{q_y\tilde{d}}{{2}}$. For a given $q_{y}$, each block, which is defined
as $h_{+}(q_{y})$ or $h_{-}(q_{y})$, represents a one-dimensional hybridised
$s$-$p$ model\cite{Li2013,Zheng2014,Zhang2014}, as shown in Fig.~\ref{fig4}, and may support edge states with
energies $\epsilon (q_{y})$ for certain values of $q_{y}$.

\begin{figure}[tbph]
\centering
\includegraphics[width=1\linewidth]{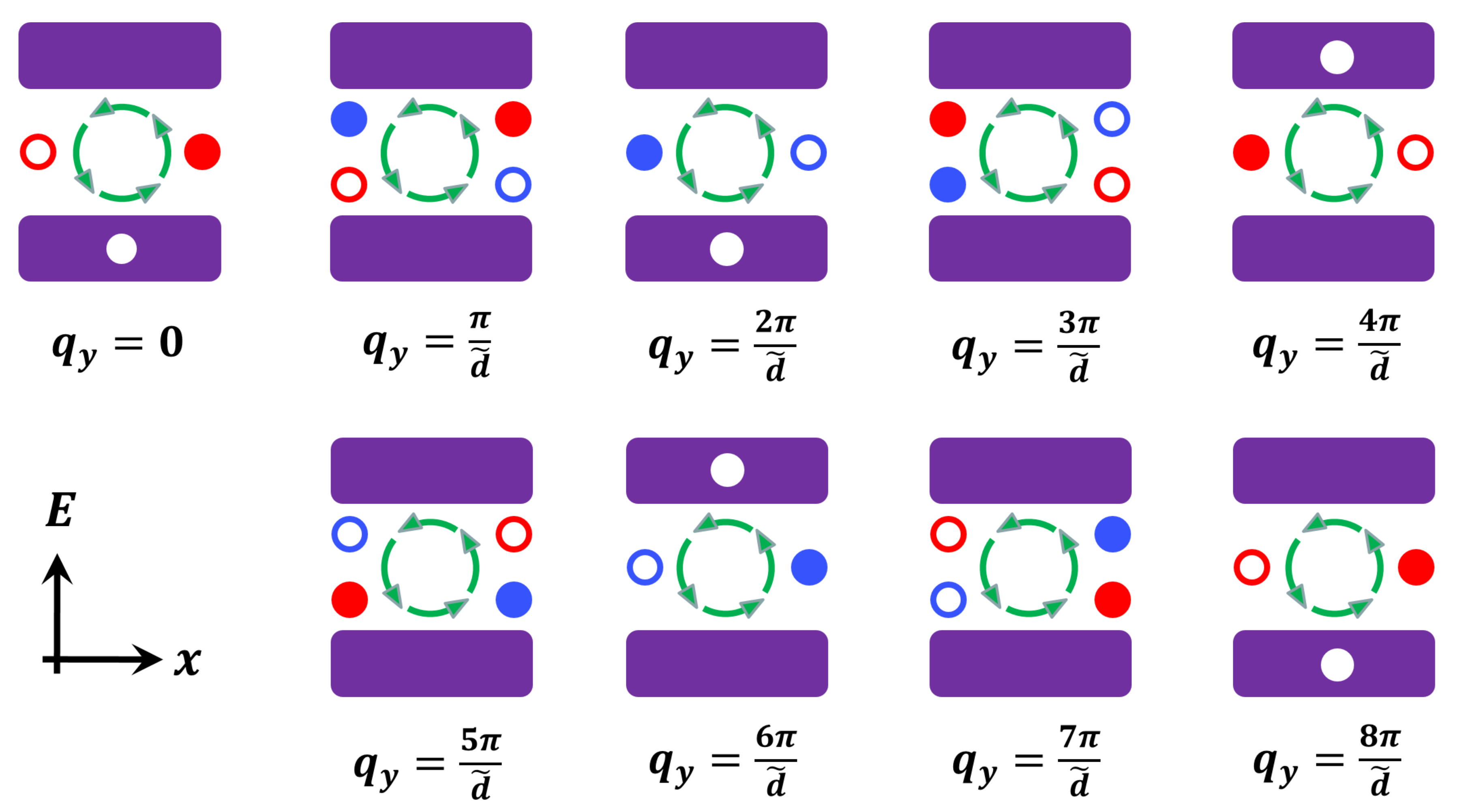}
\caption{Schematics of the
adiabatic quantum pumping with A-B-A boundaries. The purple rectangles represent the
filled bulk states and the white circle means a hole therein. The red and blue solid (open) circles represent the filled
(unfilled) edge states with the nonsymmorphic parities, $+$ and $-$, respectively. Green circular arrows represent the anticlockwise direction of the quantum pumping.}
\label{pump}
\end{figure}

Either $h_+(q_y)$ or $h_-(q_y)$ is a dimensional-reduction of the Hamiltonian of an ordinary Chern insulator. As $h_\pm(q_y)=h_\pm(q_y+4\pi/\tilde{d})$, one sees that the periods of $h_+(q_y)$ and $h_-(q_y)$ double the one
of $h(q_y)$. Moreover, $h_+(q_y+{2}\pi/\tilde{d})=h_-(q_y)$ is
satisfied, as the $A-B-A$ edges respects the nonsymmorphic crystalline
symmetry. Thus the edge BZ can be regarded as the one folded from $h_+(q_y)$
or $h_-(q_y)$, and the eigenstates at the edges also show up in pairs,
unless some of them merge into the bulk spectrum, as shown in Fig.~\ref{fig4}%
. For comparison, we also show the edge states along the $A-A-A$ edge, which
are the same as those of a conventional Chern insulator, since such a choice
of edge does not respect the nonsymmorphic symmetry. For the $A-B-A$ edge,
the number of edge states depends on $q_y$, and there might be one or two
edge states along each edge. For certain values of value of $q_y$, both
Chern insulators corresponding to $h_+(q_y)$ and $h_-(q_y)$, respectively,
have edge states, whereas for other values of $q_y$, only one of them
contributes an edge state. Such feature of the edge states
also has interesting consequence in the quantum pumping, which has an extra degree of freedom associated with the nonsymmorphic symmetry. As shown in Fig.~\ref{pump}, an edge state with parity $+$ is initially prepared at the right side of the system when $q_y=0$. Increasing $q_y$, such edge state gradually emerges to the bulk and eventually shows up on the other side of the system, when $q_y=2\pi/{\tilde{d}}$. Note that this is only half of the period of either $h_+(q_y)$ or $h_-(q_y)$. Thus, neither of them contributes to a quantised charge pumping. However, the total contribution is a well quantised one. Moreover, the parity of the edge changes to $-$. When $q_y$ continuously increases from  $2\pi/{\tilde{d}}$ to $4\pi/{\tilde{d}}$, there is no charge pumping between the two edges. However, the parity at the left side of the system changes to $+$. When $q_y$ increases to $6\pi/{\tilde{d}}$ and $8\pi/{\tilde{d}}$, the right edge is occupied with the parity $-$ and $+$, respectively. From this process, one sees that the parity of nonsymmorphic symmetry emerges as an additional degree of freedom in the topological pumping.

\textbf{Breaking the nonsymmorphic symmetry}~
Whereas we have been focusing on the lattice potential that respects the
nonsymmorphic crystalline symmetry, it is useful to discuss perturbations
that may break such symmetry. To concretise our discussions, we focus on {%
\begin{equation}
V^{\prime }=\alpha \sin \frac{\pi (x-y)}{d}\sin \frac{\pi (x+y)}{d}.
\end{equation}%
A small $V^{\prime }$ opens up a finite gap $\delta $ at the zone boundary
between the lowest(highest) two bands. Similar to other symmetry protected
topological states, if $\delta $ is much smaller than relevant energy scales
in the unperturbed system, it is expected that such perturbation will not
change the previously obtained results qualitatively. For instance, when $%
\delta \ll \Delta $, the exponentially small mixing with the highest two
bands is negligible, and one is allowed to focus on the Hilbert space
spanned by the lowest two bands. The new eigenstates thus correspond to a
unitary transformation of the unperturbed ones, which is denoted as $U$.
Correspondingly, the non-abelian Berry curvature, which is gauge dependent,
becomes $\mathbf{F}\rightarrow U^{\dagger }\mathbf{F}U$. Nevertheless, the
first Chern number is gauge independent, as the trace of the tensor remains
unchanged. Thus, the quantum charge pumping remains unaffected. It is worth
mentioning that, when $\delta $ becomes finite, it is certainly allowed to
use abelian Berry phase to compute the Chern number of each individual band
separately. Interestingly, $C_{1}+C_{2}=C$, where $C_{1}$ and $C_{2}$ are the
Chern numbers of the first and second bands computed from abelian Berry
curvature, respectively. Such an identity can be easily seen from $
\mathbf{F}_{1}+\mathbf{F}_{2}=\text{tr}\left( \mathbf{F}\right)$,
where $\mathbf{F}_{1(2)}$ is the abelian Berry curvature for the first (second) band.
We further consider tuning $\alpha$ in $V^{\prime }$, so that $%
\alpha=0$ corresponds to a transition point, at which $\delta $ vanishes.
On either side of the transition point, $C_{1}$ and $C_{2}$ are computed,
and we find out that $C_{1}+C_{2}$ is conserved, since $
\mathbf{F}_{1}+\mathbf{F}_{2}=\text{tr}\left( \mathbf{F}\right)$ is satisfied on both sides of the transition point. }

For a small $\delta$, the qualitative result of Wilson line remains
unchanged, except for a quantitative difference that it is no longer a step
function, but changes smoothly across the zone boundary, similar to the
one-dimensional system we considered recently \cite{Zhou2016}. We have also
verified that the results of edge states also remain qualitatively the same,
as shown in the supplementary material. One could also retain the nonsymmorphic crystalline symmetry and break the symmetry $M_x$ or $M_y$. The nodal boundary then evolves to a nodal line in the first BZ, which depends on the microscopic parameters of the system(supplementary material). Our system thus provides physicists a highly controllable platform to create and manipulate the nodal lines in the momentum space.

\textbf{Conclusion}~
We have discussed how to fold the energy bands of a Chern insulator to
obtain a new type of topological matter, whose BZ boundary is nodal. Non-abelian Berry curvature and nonsymmorphic symmetries become crucial in such system, and give rise to a number of unique properties in both the bulk and the edge that are distinct from ordinary Chern insulators. We hope that our work will stimulate more studies of non-abelian quantum topological matters and crystalline symmetries in ultracold atoms.

\bibliography{ref}
\bibliographystyle{ScienceAdvances}

\onecolumngrid

\section*{Supplementary Material}

\subsection{Band folding by a transformation in real space}
\begin{figure}[tbhp]
\centering
\includegraphics[width=1\linewidth]{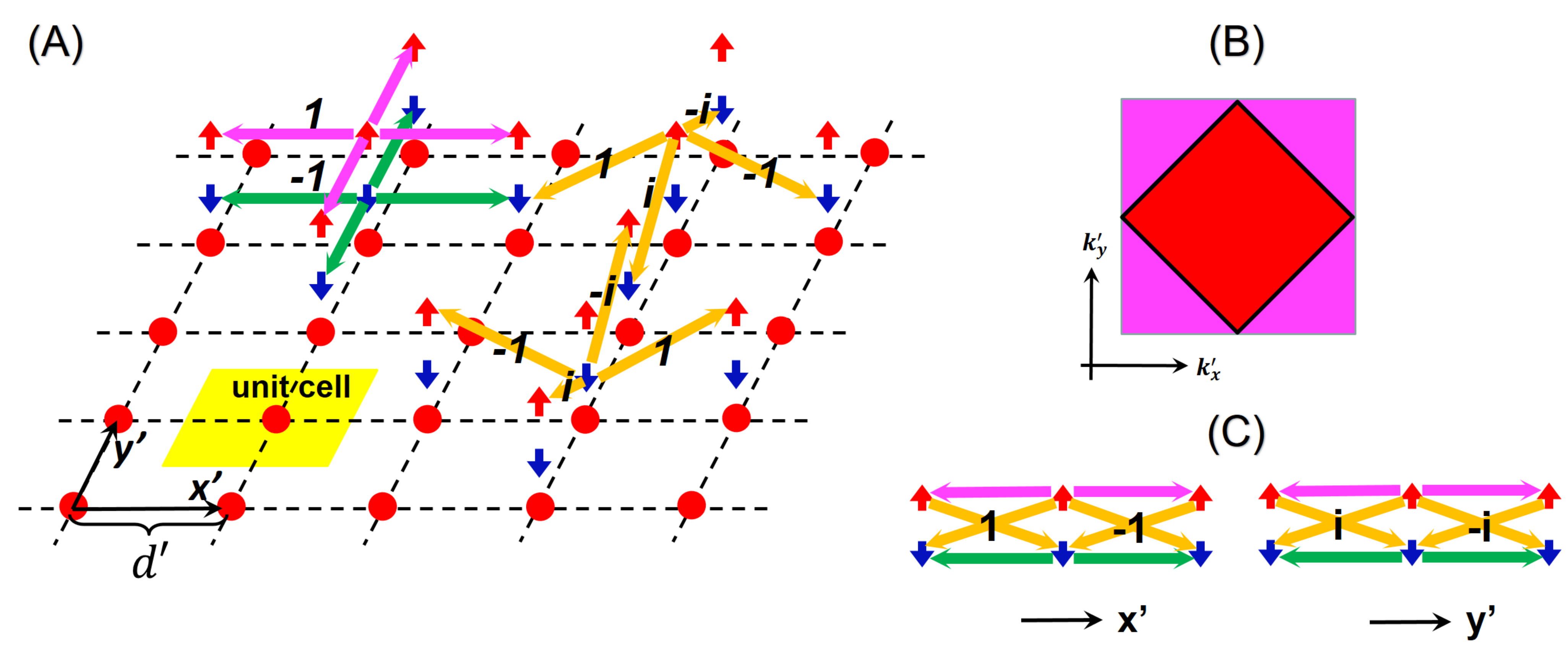}
\caption{Folding in real space. ({\bf A}) Schematics of tunnelings. After the transformation, $b^\dag_{{\bf m}\downarrow}\rightarrow -b^\dag_{{\bf m}\downarrow}$, A and B
sublattice sites in Fig.~\ref{fig1}A of the main text become equivalent, and are denoted by red spheres uniformly. Green
arrows represent intra-spin tunnellings with inverse sign of the pink
arrows. A reduced unit cell has been highlighted by yellow colour. Other
notations are the same to those of Fig.~\ref{fig1}A in the main text. ({\bf B}) Red region
is the 1st BZ of the original system, while addition of the pink
one yields the enlarged 1st BZ of the transformed system. ({\bf C}) Schematic of the tunnelling along $\hat{x}'$ and $\hat{y}'$ directions, respectively. Changing the sign of the local basis in the odd lattice sites of spin-up atom, the model in the main text is recovered.}
\label{fold}
\end{figure}

Under the transformation in real space, say $b^\dag_{{\bf m}\downarrow}\rightarrow -b^\dag_{{\bf m}\downarrow}$, the
tight-binding Hamiltonian \ref{rtb} in the main text becomes
\begin{equation}
\hat{H}=\hat{T}_{\uparrow }+\hat{T}_{\downarrow }+\hat{T}_{+}+\hat{T}_{-}+%
\hat{V},  \label{tbf}
\end{equation}%
where
\begin{eqnarray}
\hat{T}_{\uparrow } &=&t\sum_{\mathbf{m}}\Big(\hat{c}_{\mathbf{m}\uparrow
}^{\dagger }\hat{c}_{\mathbf{m}+d^{\prime }\hat{x}^{\prime },\uparrow }+\hat{%
c}_{\mathbf{m}\uparrow }^{\dagger }\hat{c}_{\mathbf{m}+d^{\prime }\hat{y}%
^{\prime },\uparrow }+\text{H.c.}\Big), \\
\hat{T}_{\downarrow } &=&-t\sum_{\mathbf{m}}\Big(\hat{c}_{\mathbf{m}%
\downarrow }^{\dagger }\hat{c}_{\mathbf{m}+d^{\prime }\hat{x}^{\prime
},\downarrow }+\hat{c}_{\mathbf{m}\downarrow }^{\dagger }\hat{c}_{\mathbf{m}%
+d^{\prime }\hat{y}^{\prime },\downarrow }+\text{H.c.}\Big), \\
\hat{T}_{+} &=&-t^{\prime }\sum_{\mathbf{m}}\Big(\hat{c}_{\mathbf{m}\uparrow
}^{\dagger }\hat{c}_{\mathbf{m}+d^{\prime }\hat{x}^{\prime },\downarrow }-%
\hat{c}_{\mathbf{m}\uparrow }^{\dagger }\hat{c}_{\mathbf{m}-d^{\prime }\hat{x%
}^{\prime },\downarrow }-i\hat{c}_{\mathbf{m}\uparrow }^{\dagger }\hat{c}_{%
\mathbf{m}+d^{\prime }\hat{y}^{\prime },\downarrow }+i\hat{c}_{\mathbf{m}%
\uparrow }^{\dagger }\hat{c}_{\mathbf{m}-d^{\prime }\hat{y}^{\prime
},\downarrow }+\text{H.c.}\Big), \\
\hat{T}_{-} &=&t^{\prime }\sum_{\mathbf{m}}\Big(\hat{c}_{\mathbf{m}%
\downarrow }^{\dagger }\hat{c}_{\mathbf{m}+d^{\prime }\hat{x}^{\prime
},\uparrow }-\hat{c}_{\mathbf{m}\downarrow }^{\dagger }\hat{c}_{\mathbf{m}%
-d^{\prime }\hat{x}^{\prime },\uparrow }+i\hat{c}_{\mathbf{m}\downarrow
}^{\dagger }\hat{c}_{\mathbf{m}+d^{\prime }\hat{y}^{\prime },\uparrow }-i%
\hat{c}_{\mathbf{m}\downarrow }^{\dagger }\hat{c}_{\mathbf{m}-d^{\prime }%
\hat{y}^{\prime },\uparrow }+\text{H.c.}\Big), \\
\hat{V} &=&m_{z}\sum_{\mathbf{m}}\Big(\hat{c}_{\mathbf{m}\uparrow }^{\dagger
}\hat{c}_{\mathbf{m}\uparrow }-\hat{c}_{\mathbf{m}\downarrow }^{\dagger }%
\hat{c}_{\mathbf{m}\downarrow }\Big).
\end{eqnarray}%
As shown in Fig.~\ref{fold}A, the unit cell is half of that of
the original system in Fig.~\ref{fig1}A of the main text, because A and B sublattices are equivalent right
now, and we denote the annihilation (creation) operator as $\hat{c}_{\bf m}^{(\dag)}$ with ${\bf m}$ being the coordinate of the lattice site. For convenience, we rotate the coordinate axes by $-\pi/4$, and define $x'=(x-y)/\sqrt{2}$, $y'=(x+y)/\sqrt{2}$, and the lattice spacing $d^{\prime }=d/\sqrt{2}$, as shown in Fig.~\ref{fold}A. The corresponding Hamiltonian in
momentum space is $\hat{H}=\sum_{\mathbf{k}}\Psi_{\mathbf{k}%
}^{\dag }H(\mathbf{k})\Psi_{\mathbf{k}}$, where $\Psi_{\mathbf{k}%
}^{\dag }=\left( \hat{c}_{\mathbf{k}\uparrow }^{\dagger },\hat{c}_{\mathbf{k}%
\downarrow }^{\dagger }\right) $, and
\begin{equation}
H(\mathbf{k)}=\left(
\begin{array}{cc}
m_{z}+\Omega _{\mathbf{k}} & -N_{\mathbf{k}} \\
-N_{\mathbf{k}}^{\ast } & -m_{z}-\Omega _{\mathbf{k}}%
\end{array}%
\right) ,  \label{ftb}
\end{equation}%
is a $2\times2$ matrix. Here $\Omega _{\mathbf{k}}=2t\left( {\cos k'_{x}d}^{\prime }+\cos %
k_{y}'d^{\prime }\right) $ and $N_{\mathbf{k}}=2it^{\prime }\left(\sin
k_{x}'d^{\prime }-i\sin k_{y}'d^{\prime }\right) $. Eq.~\ref{ftb} is just one block, $H_+({\bf k})$, of the
Hamiltonian Eq.~\ref{block} in the main text, and the 1st
Brillouin zone (BZ) thus doubles the original one (Fig.~\ref{fold}B). Likewise, if we use another transformation, $b^\dag_{{\bf m}\uparrow}\rightarrow -b^\dag_{{\bf m}\uparrow}$, in real space, we can get another block, $H_-({\bf k})$, of the
Hamiltonian Eq.~\ref{block} in the main text.

\subsection*{An alternative equivalent model}
\begin{figure}[tbhp]
\centering
    \includegraphics[width=1\textwidth]{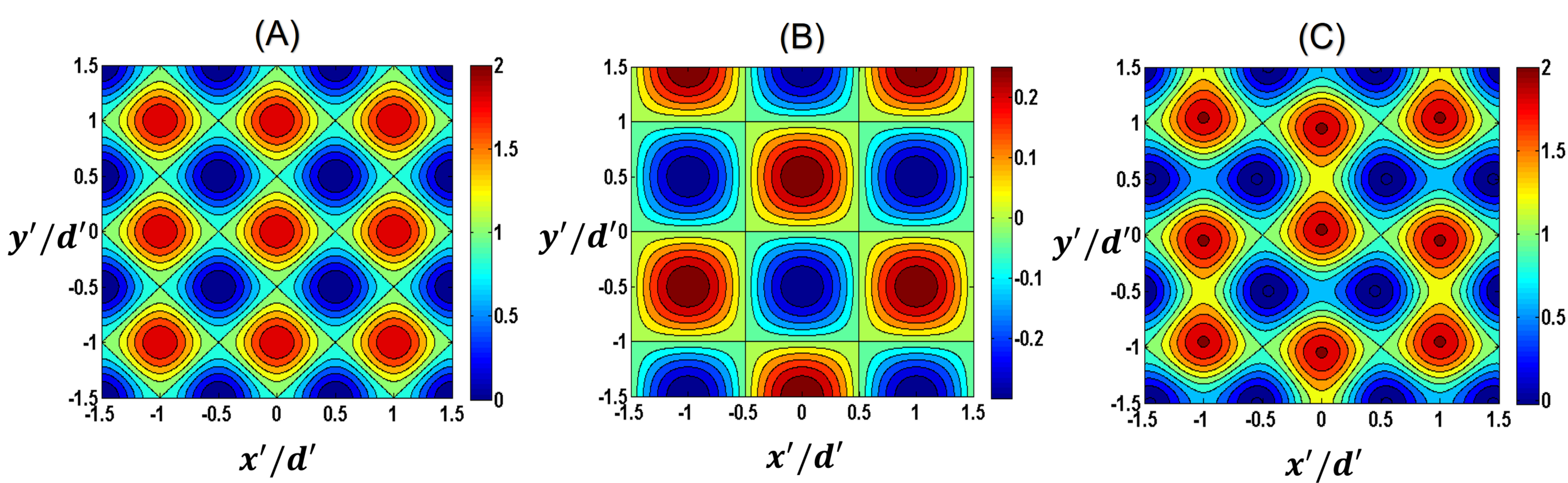}
\caption{The potentials in Eq.~\ref{pot}. ({\bf A}) The spin-independent square lattice potential, $V_0(\mathbf{r})$, with lattice spacing $d'$, where $V=E_R$, $E_R=\hbar^2 k_0^2/2M, k_0=\pi/d'$, and $M$ is the mass of the atoms. ({\bf B}) The spin-dependent potential, $V_s(\mathbf{r})$, with lattice spacing $d$, where $V_s=0.3E_R$. ({\bf C}) The two dimensional double-well lattice potential, induced by both $V_0(\mathbf{r})$ and $V_s(\mathbf{r})$, where $V$ and $V_s$ are taken the same values as those in (A) and (B). }
\label{alt}
\end{figure}

By a unitary transformation, $\sigma _{x}\rightarrow \sigma _{z}$, $\sigma
_{z}\rightarrow \sigma _{x}$, of the Hamiltonian (Eq.~\ref{rH}) in the main text, we get an alternative equivalent Hamiltonian as
\begin{equation}
V(\mathbf{r})=V_{0}(\mathbf{r})\sigma _{0}+V_{s}(\mathbf{r})\sigma
_{z}+B_{y}(\mathbf{r})\sigma _{y}+B_{x}\sigma _{x},
\label{pot}
\end{equation}%
where Pauli matrices can be regarded as real spins. Using the rotated axis by $-45^\circ$ as $x'=(x-y)/\sqrt{2}$ and $y'=(x+y)/\sqrt{2}$, $V_{0}(\mathbf{r}%
)=V\left( \cos ^{2}\frac{\pi x'}{d'}+\cos ^{2}\frac{\pi y'}{d'}\right) $ is a
spin-independent potential yielding square lattices with lattice spacing $d'=d/\sqrt{2}$, and
$V_{s}(\mathbf{r})=V_{s}\cos \frac{\pi x'}{d'}\sin \frac{\pi y'}{d'}$ is a
spin-dependent potential with lattice spacing $d$, as shown in Figs.\ref{alt}A and \ref{alt}B. $V$ and $V_s$ are the corresponding potential strengths. The first two
terms in Eq.~\ref{pot} yield a spin-dependent two-dimensional double-well
lattices, along $x'$ direction of which is an effective one-dimensional
double-well lattice, as shown in Fig.~\ref{alt}C. $B_{x}$ and $B_{y}(%
\mathbf{r})=B\sin \frac{\pi x'}{d}\cos \frac{\pi y'}{d}$ are the space-independent and dependent effective magnetic
fields along $x$ and $y$ directions, respectively. $B$ is the strength of the magnetic field. If we make $B=V_s$, the model is just the one we used in the main text.

\subsection*{Measuring the eigenvalue of the nonsymmorphic operators}
The nonsymmorphic operator $G_{x+y}$ shifts the particle along both the $x$ and $y$ directions for half of the lattice spacing, and meanwhile rotates the spin about the $z$ axis for $\pi$. Physically, the eigenvalue of $G_{x+y}$ is the phase difference of the wavefunction at ${\bf R}$ and ${\bf R}'$, where ${\bf R}$ includes both the real space coordinate and the spin, and ${\bf R}\rightarrow {\bf R}'$ under $G_{x+y}$. The standard technique for measuring the single particle correlation function can then be directly generalised to our system\cite{Clade2009,Navon2015}.  One simply needs to  add a $\pi$ pulse in the spin space in addition to shifting a copy of the sample by half of the lattice spacing.

If one is only interested in the parity of nonsymmorphic symmetry $\lambda=\pm$, a simple Ramsey interferometry can be used. Preparing an initial state  $
\Psi _{i}(\mathbf{r})=[\Psi _{1{\bf k }}(\mathbf{r%
})+\Psi _{2{\bf k}}(\mathbf{r})]/\sqrt{2}$, its evolution with time $t$
yields $\Psi _{f}(\mathbf{r})=[\Psi _{1{\bf k}}(%
\mathbf{r})+e^{-i\omega t}\Psi _{2{\bf k}}(\mathbf{r})]/\sqrt{2}$, where
$\hbar \omega $ is the energy difference between the lowest two bands at momentum ${\bf k}$. Applying a
$\pi /2$ pulse and $\Psi _{1{\bf k}}(\mathbf{r})\rightarrow [\Psi _{1{\bf k}}(\mathbf{r})+\Psi _{2{\bf k}}(\mathbf{r})]/\sqrt{2}$, $\Psi _{2{\bf k}}(\mathbf{r})\rightarrow [\Psi _{1{\bf k}}(\mathbf{r})-\Psi _{2{\bf k}}(\mathbf{r})]/\sqrt{2}$, one detects the probability of occupying the
lowest band as $P(t)=\cos ^{2}(\omega t/2)$. Alternatively, applying the nonsymmorphic operator $G_{x-y}$ to $\Psi _{i}(\mathbf{r})$, one obtains, $\Psi'_{i}(\mathbf{r})= ie^{i\left(
k_{x}-k_{y}\right) d/2}[\Psi _{1{\bf k }}(\mathbf{r%
})-\Psi _{2{\bf k}}(\mathbf{r})]/\sqrt{2}$. Such nonsymmorphic operation is directly available by simultaneously shifting the lattice, i.e., changing the relative phase of the lasers,  and rotating the spin. One then repeats the previous steps, the probability becomes $P'(t)=\sin
^{2}(\omega t/2)$. The phase shift between $P'(t)$ and $P(t)$ is a direct consequence of the difference $\lambda$ between the lowest two bands.

\subsection{Breaking the nonsymmorphic symmetry}
\begin{figure}[tbhp]
\centering
\includegraphics[width=1\textwidth]{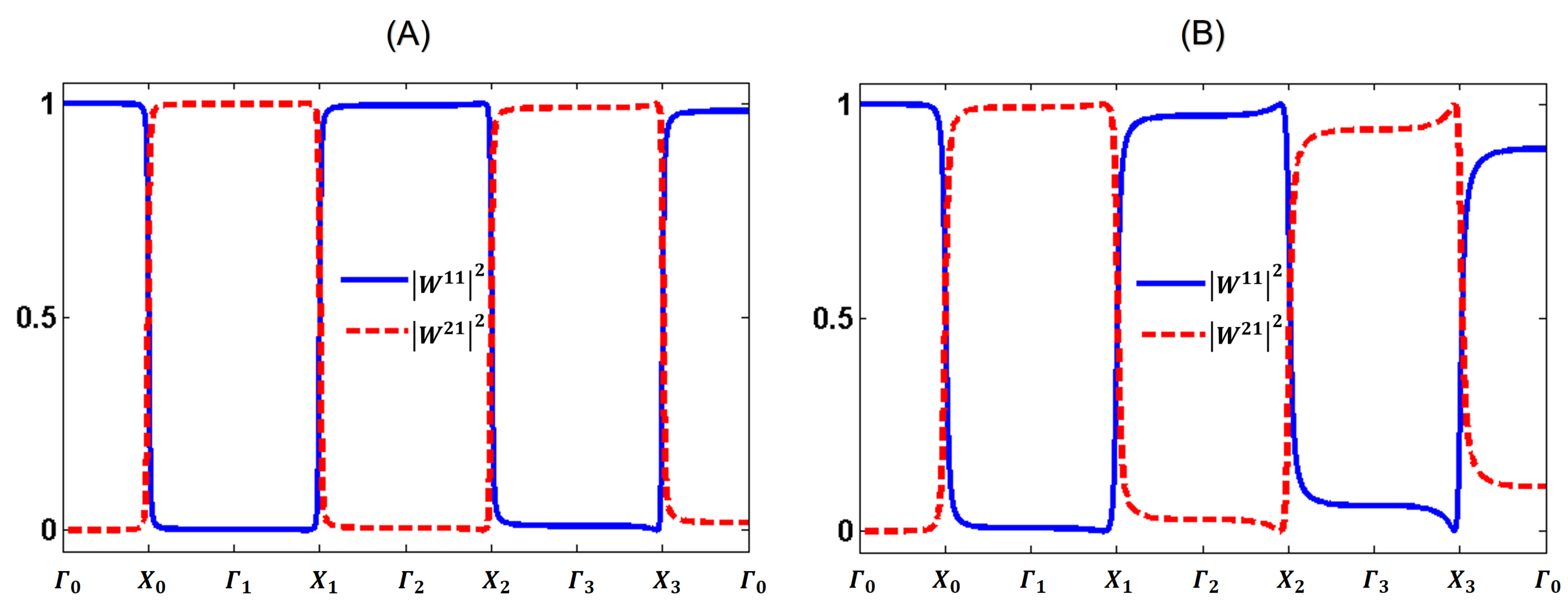}
\caption{Squared amplitudes of the two elements, $W^{11}$ and $W^{21}$, of
the Wilson loop denoted in Fig.~\protect\ref{fig3}A in the main text after
breaking the nonsymmorphic symmetry with the energy offset ({\bf A}) $\delta=0.2t$ and ({\bf B}) $\protect\delta=0.5t$. The other parameters of the
tight-binding model are $t^{\prime }=t, m_z=-3t$. }
\label{bwilson}
\end{figure}

\begin{figure}[tbhp]
\centering
\includegraphics[width=1\textwidth]{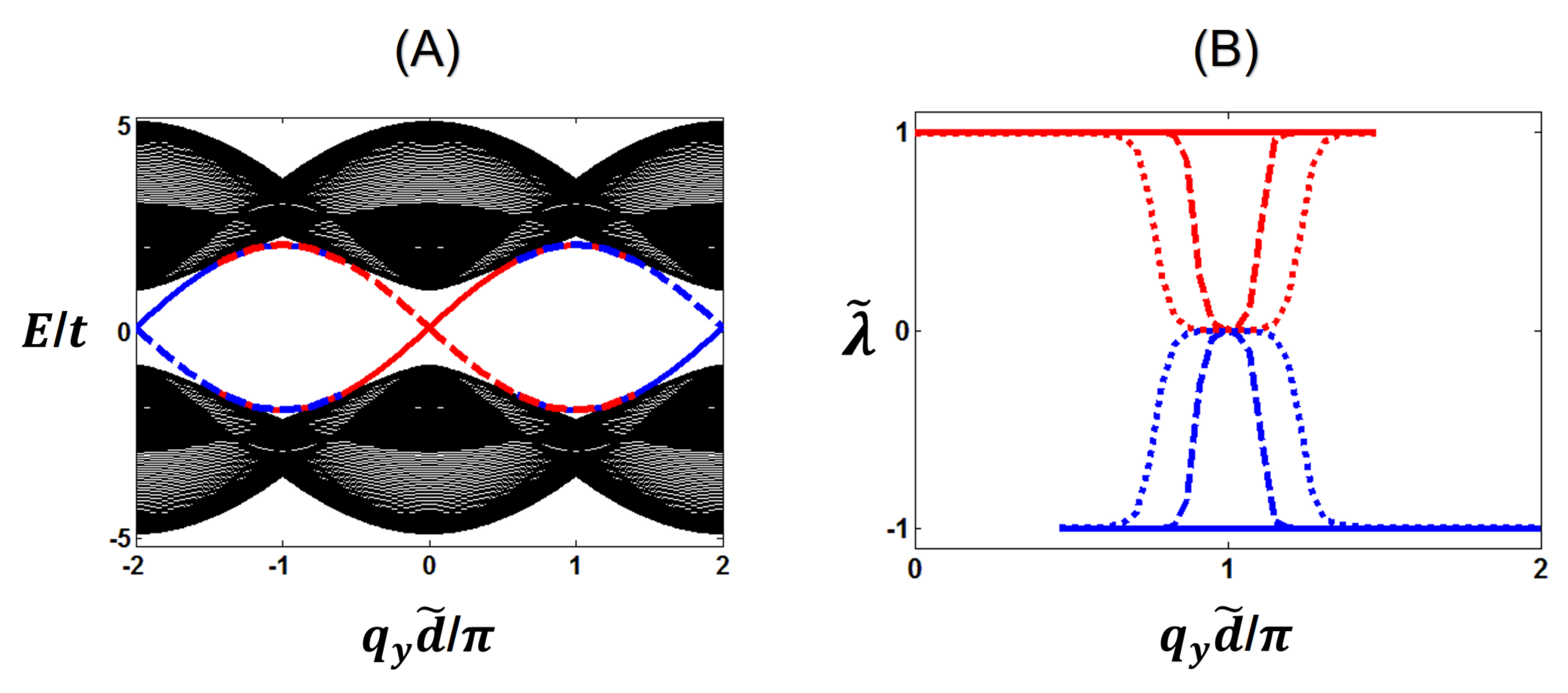}
\caption{Breaking nonsymmorphic symmetry. ({\bf A}) Energy spectra with  A-B-A boundaries after breaking the nonsymmorphic symmetry by $\delta=0.2t$. The solid and dashed lines represent the edge states on the right and left boundaries, respectively. The nondegenerate parts with single red or blue colors represent $\tilde{\lambda} \simeq 1$ or $-1$, respectively, while for the overlapping parts of the red and blue curves the parity is not well defined. ({\bf B}) The parity $\tilde{\lambda}$ of the $E/t>0$ edge states for $q_y>0$ part with $\delta=0$ (solid), $0.001t$ (dashed), and $0.1t$ (dotted). That of $q_y<0$ is symmetric with respect to $q_y=0$. The curves terminate when the edge states merge into the bulk. In the vicinity of ${q}_y\tilde{d}=\pm\pi$, edge states are strongly coupled to the bulk and those at the other side of the system such that the parity is not well defined, corresponding to the overlapping parts of the red and blue curves of edge states in (A). When ${q}_y\tilde{d}=0$ and $\pm2\pi$, edge states are separated from the bulk by a finite gap and the parity is thus well conserved. Other tight-binding model parameters used here are $t^{\prime}=t, m_z=-t$.}
\label{bedge}
\end{figure}

By introducing an additional potential,
\begin{equation}
V^{\prime }=\alpha \sin \frac{\pi (x-y)}{d}\sin \frac{\pi (x+y)}{d},
\end{equation}%
an energy offset, $\delta $, between A and B sublattice sites is turned on
and thus the nonsymmorphic symmetry is broken. The corresponding
tight-binding model is similar to Eq.~\ref{tb} with only the additional
energy offset term in the diagonal elements. It reads
\begin{equation}
H(\mathbf{k)}=\left(
\begin{array}{cccc}
m_{z} & \Omega _{\mathbf{k}}e^{i\theta _{\mathbf{k}}} & 0 & N_{\mathbf{k}%
}e^{i\theta _{\mathbf{k}}} \\
\Omega _{\mathbf{k}}e^{-i\theta _{\mathbf{k}}} & m_{z}+\delta & -N_{\mathbf{k%
}}e^{-i\theta _{\mathbf{k}}} & 0 \\
0 & -N_{\mathbf{k}}^{\ast }e^{i\theta _{\mathbf{k}}} & -m_{z} & \Omega _{%
\mathbf{k}}e^{i\theta _{\mathbf{k}}} \\
N_{\mathbf{k}}^{\ast }e^{-i\theta _{\mathbf{k}}} & 0 & \Omega _{\mathbf{k}%
}e^{-i\theta _{\mathbf{k}}} & -m_{z}+\delta%
\end{array}%
\right) ,
\end{equation}%
where $\Omega _{\mathbf{k}}=2t\left[ {\cos \frac{\left( k_{x}-k_{y}\right) d%
}{2}+\cos }\frac{\left( k_{x}+k_{y}\right) d}{2}\right] ,N_{\mathbf{k}%
}=2it^{\prime }\left[ {\sin \frac{\left( k_{x}-k_{y}\right) d}{2}-i\sin }%
\frac{\left( k_{x}+k_{y}\right) d}{2}\right] $, and $\theta _{\mathbf{k}}=%
\frac{\left( k_{y}-k_{x}\right) d}{2}$. Using this model, as in the main
text, we can also calculate the Wilson loop, which is not a step function
anymore but qualitatively the same to that preserving the nonsymmorphic
symmetry, as shown in Fig.~\ref{bwilson}. The energy spectra with A-B-A boundaries are also plotted in Fig.~\ref{bedge}A, which are also qualitatively the same to those of the system without
breaking the nonsymmorphic symmetry. Though the bulk states can no longer be labeled by the parity of the nonsymmorphic symmetry, such label still works for the edge state, if its energy is well separated from the bulk by a finite gap such that the perturbation induced mixing between $\lambda=+$ and $\lambda=-$ is exponentially small for the edge state. Fig.~\ref{bedge}B shows  $\tilde{\lambda}=\langle \phi_e | G_e |\phi_e\rangle/(i e^{ip_y \tilde{d}/2})$, where $|\phi_e\rangle$ is the edge state, and $G_e$ is the nonsymmorphic operator along the A-B-A edge,  as a function of the momentum $q_y$. It is clear that $\tilde{\lambda}$ remains to be 1 or $-1$, with exponentially small corrections, unless the edge state begins to merge into the bulk states, or coupled to edge states with the opposite parity in a small system.

\subsection{Evolution from nodal boundary to nodal line}
In general, the Hamiltonian Eq.~\ref{rH} in the main text can be anisotropic along $x+y$ and $x-y$ directions, of which the potentials become
\begin{eqnarray}
V(\mathbf{r})&=&V_x\cos ^{2}\frac{\pi (x-y) }{d}+V_y\cos ^{2}\frac{\pi (x+y) }{d}, \\
\Omega _{x}(\mathbf{r})&=&\Omega_x \cos \frac{\pi
(x-y) }{d}\sin \frac{\pi (x+y) }{d}, \\
\Omega _{y}(\mathbf{r})&=&-\Omega_y \sin \frac{\pi (x-y) }{d'}\cos \frac{\pi
(x+y) }{d}.
\end{eqnarray}
In the following, for convenience, we again  use the rotated axes, $x'=(x-y)/\sqrt{2}$ and $y'=(x+y)/\sqrt{2}$, and the lattice spacing $d^{\prime }=d/\sqrt{2}$.

\begin{figure}[tbhp]
\centering
\includegraphics[width=1\textwidth]{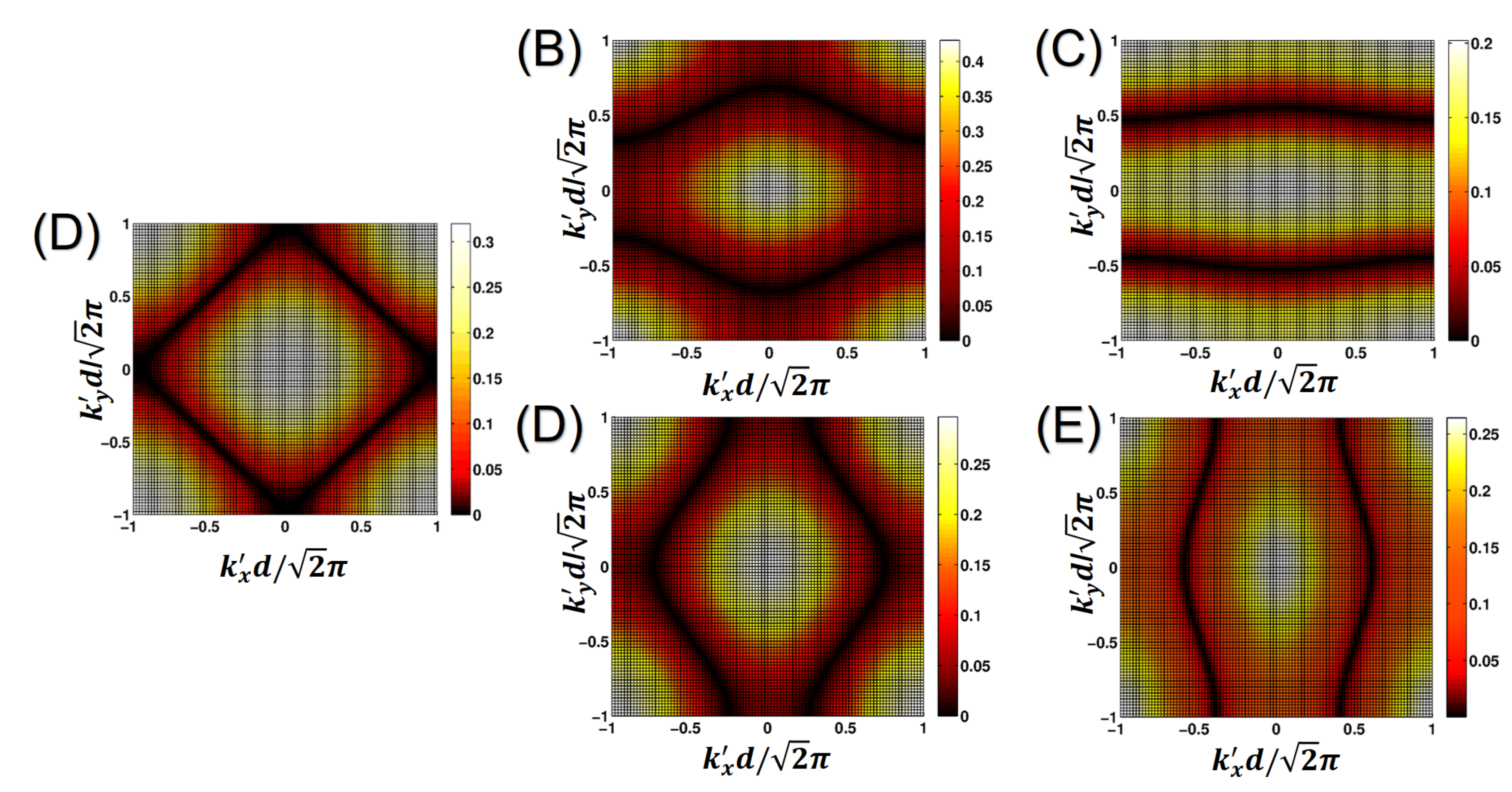}
\caption{Evolution of the nodal line. The contour plots are the energy difference of the lowest two bands by plane-wave expansions, and the black curves are the nodal lines. ({\bf A}) The isotropic case with $(V_x,V_y)=(10,10)E_R$ and $(\Omega_x,\Omega_y)=(5,5)E_R$. ({\bf B,C}) The anisotropy induced by the lattice potentials with $(V_x,V_y)=(10,8)E_R$ and $(\Omega_x,\Omega_y)=(5,5)E_R$, and $(V_x,V_y)=(10,0)E_R$ and $(\Omega_x,\Omega_y)=(5,5)E_R$, respectively. ({\bf D,E}) The anisotropy induced by the Raman potentials with $(V_x,V_y)=(10,10)E_R$ and $(\Omega_x,\Omega_y)=(5,4)E_R$, and $(V_x,V_y)=(10,10)E_R$ and $(\Omega_x,\Omega_y)=(5,0)E_R$, respectively. All are with $m_z=-E_R$. $E_R = \hbar^2k^2_0/2M$ is the recoil energy with $k_0 = \pi/d'$ and M being the mass of the atoms. }
\label{nodalline}
\end{figure}

In the main text, we have proved that the nodal boundary is protected by both the nonsymmorphic symmetry, $G_{x\pm y}$, and the symmetries $M_x$ and $M_y$. When  $M_x$ and $M_y$ are broken, the nodal  BZ boundary turns to a nodal line.  As shown in Fig.~\ref{nodalline}, the nodal line deforms when changing the anisotropy of either the lattice potentials, $V_x\neq V_y$ (Fig.~\ref{nodalline}A,B,C), or the Raman potentials, $\Omega_{x}\neq \Omega_{y}$ (Fig.~\ref{nodalline}A,D,E). The nodal line always exists because the nonsymmorphic symmetry, $G_{x\pm y}$, is preserved, though $M_x$ and $M_y$ are broken. Interestingly, as shown in Fig.~\ref{nodalline}, the nodal line always get through the four $X$ points, i.e. ${\bf k}=(0,\pm \pi/d),(\pm\pi/d,0)$, no matter how anisotropic the system is. Such band crossing points are protected by the inversion symmetry, $I$: $x'\rightarrow d'-x', y'\rightarrow -y'$, which, combined with the nonsymmorphic symmetry, protects the four nodal $X$ points in BZ. To see this, one notices the relation,
\begin{equation}
IG_{x-y}=G_{x-y}IT(d,-d),
\end{equation}
where $T(d,-d)$ is the translation operator, which translates the state by $d$ and $-d$ along $x$ and $y$ directions, respectively. For Bloch states, it becomes
\begin{equation}
IG_{x-y}=G_{x-y}Ie^{i(k_x-k_y)d},
\end{equation}
which shows, at $X$ points, the inversion operator and the nonsymmorphic operator are anticommutative, so the degeneracy is guaranteed.

\end{document}